# Seasonal Predictability of Lightning over the Global Hotspot Regions

Chandrima Mallick[1,2], Anupam Hazra[1]*, Subodh K. Saha[1], Hemantkumar S. Chaudhari[1], Samir Pokhrel[1], Mahen Konwar[1], Ushnanshu Dutta[1,2], Greeshma M. Mohan[1], and K. Gayatri Vani[1]

[1]Indian Institute of Tropical Meteorology, Ministry of Earth Sciences, Pune, India.

[2]Department of Atmospheric and Space Sciences, Savitribai Phule Pune University, Pune, India.

Corresponding author: Anupam Hazra (hazra@tropmet.res.in)

**Key Points:**

- Possibility of reliable seasonal forecasting of lightning over global hotspots is demonstrated.

- Lightning flashes are found to be tied with slowly varying remote global predictors (e.g., El Niño and Southern Oscillation, Atlantic Multi-decadal Oscillation).

- Multiple regression analysis demonstrates that seasonal lightning and associated rainfall is predictable during pre-monsoon and monsoon.

**Abstract**

Skillful seasonal prediction of lightning is crucial over several global hotspot regions, as it causes severe damages to infrastructures and losses of human life. While major emphasis has been given for predicting rainfall, prediction of lightning in one season advance remained uncommon, owing to the nature of problem, which is short-lived local phenomenon. Here we show that on seasonal time scale, lightning over the major global hot-spot regions is strongly tied with slowly varying global predictors (e.g., El Niño and Southern Oscillation). Moreover, the sub-seasonal variance of lightning is highly correlated with global predictors, suggesting a seminal role played by the global climate mode in shaping the local land-atmosphere interactions, which eventually affects seasonal lightning variability. It is shown that seasonal predictability of lightning over the hotspot is comparable to that of seasonal rainfall, opens up an avenue for reliable seasonal forecasting of lightning for special awareness and preventive measures.

**Plain Language Summary**

Lightning, atmospheric hazards have an impact on the loss of human life, forest fire, health, agriculture, and economy across the globe. Investigation forecasting of lightning flashes in terms of tendency in one season advance is need of the hour. However, due to its chaotic nature, the tendency of seasonal forecasting of lightning is considered not viable as the understanding of the predictability of lightning is still incomplete. Here, we have explored the possibility of seasonal forecasting of lightning activity and provide a scientific basis as the lightning flashes are found to be tied with slowly varying remote forcing forces (e.g., El Niño and Southern Oscillation, ENSO or other global predictors). Correlation of flash count with different indices (Nino, Pacific Decadal Oscillation, North Atlantic Oscillation and Extra-tropics, etc.) demonstrate the potential of seasonal forecasting of lightning. The multiple regression analysis again enhances the skill. Therefore, the better climate models that capture crucial couplings between ocean, atmosphere, and land processes could make skillful predictions of lightning and opens up a possibility for forecasting of lightning in one season advance, which will be helpful to the policymakers.

## 1. Introduction

Lightning is one of the most powerful, all pervasive atmospheric hazards. Lightning fatalities over India (Mahapatra et al., 2018; Accidental Deaths & Suicides in India (ADSI) Report, Govt. of India) result in severe societal and economic consequences as lightning affect aviation, telecommunication, agriculture, electricity and many more sectors. Keeping in mind the lifetime injuries, physiological trauma, disabilities that are caused by lightning (e.g., Mahapatra et al., 2018; Singh and Singh, 2015), an accurate or better formulation of predicting lightning activities is of utmost significance. The major question asked here is whether the global hot spots of lightning activities across the globe are associated with the remote forcing e.g. El Niño and Southern Oscillation (ENSO), Pacific Decadal Oscillation (PDO), North Atlantic Oscillation (NAO), Atlantic Multidecadal Oscillation (AMO), and Extra Tropics (ET).

The seasonal forecasting of hail and tornado occurrence, which are the severe weather phenomena associated with deep convective system are examined over the United States of America (Wu et al., 2011; Elsner and Widen, 2014). Lightning activity is also a typical phenomenon of severe weather characterized by strong convection (van den Broeke et al., 2005), where instability, strong vertical updraft, wind shear and availability of moisture are the primary conditions. However, the seasonal prediction in extra-tropics is fundamentally different from the tropics (Charney and Shukla, 1981). The higher predictability of tropical climate than extra-tropics was laid by the scientific basis of pioneering work on Indian summer monsoon (ISM) predictability by Charney and Shukla (1981). The large heat capacity in the ocean can give the climate system a memory, which can affect the atmospheric deviations lasting for months to years and the scientific basis of prediction lies in the predictability established by the conditions of the ocean and land surface (Wang et al., 2005). The physical basis for seasonal prediction in the tropics is that the low-frequency component of variability is primarily governed by slowly varying boundary forcing (e.g., SST, soil moisture, snow cover, and more), which laid the foundation for deterministic seasonal prediction in the tropics (Charney and Shukla,1981).

Recent studies are also unrevealed that the high-frequency sub-seasonal fluctuation of ISM rainfall (ISMR), so far considered chaotic, is partly predictable as it is tied with ENSO (Saha et al., 2019, 2021, Dutta et al., 2021). In recent years, the associations of lightning and thunderstorm activities with the ENSO and various other large-scale modes of atmospheric and oceanic variability such as NAO, the Quasi-Biennial Oscillation (QBO), and the Indian Ocean Dipole (IOD) are reported (e.g. de Pablo & Soriano, 2007; Bovalo et al., 2012; Dowdy, 2016). Dowdy (2016) has demonstrated that ENSO has the strongest relationship with lightning activity and weak relationship with other modes of variability during each season. Chronis et al. (2008) have also shown the relationship of global lightning climatology in response to the ENSO cycle. Thus it is revealed that lightning and thunderstorm activity are potentially predictable for several months in advance in various regions throughout the world including the region of the tropics (Dowdy, 2016; Lopez, 2016). Munoz et al., (2016) highlighted that the predictive skill of lightning density is higher than typical values for rainfall amounts in some regions (e.g., North Western Venezuela).

Over the Indian subcontinent, lightning flashes mostly occur during the pre-monsoon and post-monsoon season due to favorable atmospheric conditions, e.g. instability in the atmosphere, strong vertical updraft, wind shear and availability of moisture (Kamra, 1985, Latham et al., 2004, Barthe et al., 2010, Barth et al., 2012, Mohan et al., 2021). The flash rate is assumed to be proportional to the maximum vertical updraft velocity, fourth power of cloud dimension (where graupel exists along with snow, ice, and cloud water), and also becomes directly proportional to the fifth power of the cloud top height (Williams, 1985). Seasonal distributions of lightning activity have been explored in several studies (Kandalgaonkar, 2005, Ranalkar and Chaudhari, 2009, Tinmaker et al., 2010). In the El-Nino (La-Nina) period of the pre-monsoon and monsoon seasons over India, there is an increase (decrease) in the flash density as compared to the Non-ENSO period (Ramesh Kumar and Kamra, 2012; Ahmad and Ghosh, 2017; Sreenath et al., 2021). In another study, Kulkarni (2015) has revealed that for rainfall prediction atmospheric electricity can be used as a proxy parameter.

Therefore, in the present endeavor, we attempt to find the potential feasibility of reliable seasonal forecasting of lightning over global hot-spots along with India as a sub-set. The following questions are attempted to address here:

(i) Is there any correlation of lightning over different global hotspots with different indices (Nino, NAO, PDO, AMO and ET)?

(ii) Are the seasonal occurrences of lightning activities over global tropics and India is associated (i.e. teleconnected) with slowly varying predictable component (e.g., SST over NINO index or any other global predictors, NAO, PDO, AMO and ET)?

(iii) Whether sub-seasonal variance (i.e., a vigor of convection) is linked with slowly varying predictable component (e.g., SST over NINO index or any other global predictors, NAO, PDO, AMO, ET and more)?

The hypothesis for the seasonal forecast can be stated as lightning flashes over global hotspot regions along with India during pre-monsoon (March-May, MAM) and monsoon (June-September, JJAS) seasons can be predicted if the mean and sub-seasonal fluctuation is found to be tied with slowly varying remote global predictors. Here, we have also tried to investigate the possible relationship of lightning with the large-scale modes of variability (i.e. ENSO, NAO, PDO, AMO, ET and more) as these phenomena are predictable at least a season well in advance.

## 2. Data and Methods

The satellite-based two lightning sensors: Lighting Imaging Sensor (LIS) and Optical Transient Detector (OTD) from NASA (Cecil et al., 2014) are used in this study from 1997 to 2013 (17 years). The detailed seasonal climatology of lightning (Fig. 1) from LIS and OTD sensors on the Tropical Rainfall Measuring Mission (TRMM) satellite (Cecil et al., 2014, Blakeslee, 2021) are presented. The daily gridded lightning flash densities generated from a combination of the LIS and OTD data (available from http://thunder.nsstc.nasa.gov) (Mach et al., 2007, Blakeslee, 2021) which has a spatial resolution of 2.5x2.5 degrees is used to elucidate the distribution of lightning strikes, its mean and variability. The year to year variation in the mean Lightning flash counts for MAM (March to May) and JJAS (June to September) seasons are evaluated over the study regions. The interannual variability of lightning flash can give insight of year to year variation of lightning on the seasonal time scale. The observational datasets of SST from HadiSST (Rayner et al., 2003) are used to understand the relation of lightning with large-scale processes. It is well known that lightning has strong relationship with cloud top temperatures (Liu et al., 2012). Therefore, the cloud top temperature (CTT) data from Modern-Era Retrospective analysis for Research and Applications version 2 (MERRA2) (Gelaro et al., 2017) are also used to find the correlation with large scale predictors (i.e., SST) over the global tropics. The rainfall data from Global Precipitation and Climatology Project (GPCP, Adler et al., 2003) is also used to calculate the correlation with lightning. The data of fatalities over India due to natural disasters are obtained from the "Accidental Deaths & Suicides in India (ADSI) Report" published by National Crime Records Bureau (NCRB) data, Govt. of India ((https://ncrb.gov.in/)

for the years 2002 to 2019 (36th to 53rd series). Merchant et al. (2019) have shown that the teleconnections are large-scale atmospheric variability patterns that result from changes in slowly varying forcing mechanisms like SST, soil moisture, etc. The climatic relationships or teleconnection with global predictors like ENSO, NAO, AMO, PDO, and ET of certain variables over large distances may provide predictability of those (Mock, 2014; Chattopadhyay et al., 2015; Borah et al., 2020). The multiple linear regression analysis is also used to present the combined indices. The combined effect of different indices (e.g. ENSO, NAO, AMO, PDO, ET) on Flashes of Lightning (FL) can be obtained by the multiple linear regression analysis technique. The multiple linear regression analysis (Alexopoulos, 2010, Saha et al., 2021) has been applied in this study as shown in the following equation (1):

$$FL = C_0 + C_1 Nino_{1+2} + C_2 Nino_3 + C_3 Nino_{3.4} + C_4 Nino_4 + C_5 NAO + C_6 AMO + C_7 PDO + C_8 ET$$

-------------- (1)

Where, $C_0$ = intercept

$C_1, C_2, \ldots, C_8$ = Regression coefficients

## 3. Results

### 3.1. Seasonal mean climatology of lightning over global hotspot regions and lightning variability of the Indian subcontinent :

The seasonal mean climatology of lightning flashes over major hotspot regions over global tropics and India are represented in Figure (1a,b) and Figure (1c,d) during MAM and JJAS respectively. Two major global lightning hotspot regions are selected over North America (Region 3: 110 °W - 71 °W; 5 °N - 43 °N) and Africa (Region 4: 8 °E - 35 °E; 12 °S - 15 °N) as shown in Figure 1a,c. Similarly, over India two more lightning hot spot regions are identified: (i) East and north-East (Region 1: 82 ºE - 94 ºE ; 20 ºN - 28 ºN) (ii) North-West (Region 2: 68 ºE - 80 ºE; 25 ºN - 38 ºN), which are marked in Figure 1b,d. These lightning global hotspots are consistent with earlier studies ( Zipser et al. 2006; Choudhury et al. 2020 and references there in). It also is important to note that although the total accidental deaths due to natural disasters (e.g., Cyclone, Flood, Heat and cold wave, Lightning, Landside (Accidental Deaths & Suicides in India (ADSI) Report, Govt. of India published by National Crime Records Bureau (NCRB) data, Govt. of India) over the Indian subcontinent decreases due to the advancement of technology and forecast system particularly cyclone, flood, heat and cold wave, but deaths due to lightning remain is increasing over the years (Fig. 1e) and percentage of death increases (red line, Fig. 1e) in recent years. The increasing trend of the percentage of death due to lightning (death due to lightning/total death by all natural disasters x 100) have emerged as a new headache of the policymakers in recent years (red line, Fig. 1e). Therefore, seasonal prediction of lightning may achieve a new importance for the effective preparedness and mitigation activities.

To understand the usefulness of the lightning prediction in seasonal time scale over India, which is a sub-set of global lightning hotspot regions, firstly the year to year variation of seasonal mean lightning flash (averaged over region 1 and region 2) is analyzed for MAM (Fig. 2a) and JJAS (Fig. 2b) seasons. It is important to note that during MAM, lightning flash density is more (less) over Region 1 (Region 2) as seen in figure 2a. On the other hand, during JJAS, Region 2 experiences more lightning as compared to Region 1 (Fig. 2b). The El-Niño and La-Nina years obtained from monsoon online of Indian Institute of Tropical Meteorology (https://mol.tropmet.res.in/monsoon-interannual-timeseries/), which is based on All-India area-weighted mean summer monsoon rainfall (AISMR) are marked as stars in red (indigo) (Fig. 2a).

The percentage of interannual variability of lightning flash count during MAM and JJAS are also presented (Figure 2c,d). The prominent interannual variation of lightning flash counts indicate the importance and possibility of seasonal forecasting of lightning (Fig. 2c,d). It is interesting to note that seasonal predictions are bound to the general trends/tendencies of the phenomena than the forecast of specific days. The tendencies of lightning for a particular season over specific regions are focused for seasonal prediction. But, we do not focus on forecasts for specific days.

**3.2: Global teleconnection of lightning over global hotspot regions and SST:**

It can be hypothesized that, the lightning flashes may be predictable one season in advance if it is tied to the slowly varying forcing (e.g., El Niño and Southern Oscillation, ENSO) and other global predictors (e.g., ET, PDO, NAO, AMO). To investigate the linkages between lightning flash density and slowly varying predictable component, the mean lightning flash count averaged over the four lightning hotspot regions over global tropics are correlated with global SST (2m Temperature) over the ocean (land) during MAM and JJAS (Figure 3,4). The correlation of mean lightning flash count and cloud top temperature (CTT) averaged Region 3 and Region 4 (as shown in figure 1a,c) with global SST are calculated during MAM and JJAS (Figure 3). The deep convection over the study regions is verified by correlating the cloud top temperature (CTT) averaged over Region 3 (Fig. 3c3,d3) and Region 4 (Fig. 3c4,d4) with SST during MAM and JJAS. Interestingly, an opposite strong correlation pattern is observed between lightning and CTT. This is because lightning flashes increases when CTT decreases (Figure not shown), which is also mentioned by previous studies (Price and Rind, 1992; Liu et al., 2012). The stronger correlation in lightning (Fig. 3a3) and CTT (Fig. 3c3) during MAM indicates the strongest relationship of large scale global predictors and seasonal mean lightning flash density.

The results reveal that lightning activity over Region 3 and Region 4 shows a strong teleconnection with global SST.

The different modes of variability and its linage with lightning flashes are examined by correlating of the area averaged lightning flash counts (averaged over four regions) with different indices for MAM and JJAS as shown in figure 3e,f. The ENSO (represented by the NINO4, NINO3.4, NINO3, NINO1+2 index), the NAO, AMO, PDO and ET are considered in the present study. We have also done the multiple regression analysis for both lightning and rainfall over selected global lightning hotspot regions (Region 1, Region 2, Region 3, and Region 4) along with all India (lon: 68 $^{o}$E - 98 $^{o}$E, lat: 8 $^{o}$N - 38 $^{o}$N) to get the combined correlation coefficient and depicted in Figure (3g,h) for MAM and JJAS. The linear multiple regression analysis demonstrates that there is a strong correlation over Region 1 (R ~ 0.74), Region 2 (R ~ 0.42), Region 3 (R ~ 0.72) and Region 4 (R ~ 0.72) along with all India (R ~ 0.55) during MAM for lightning. Interestingly, the correlation value for lightning (R ~ 0.74) in MAM over Region 1 is greater than rainfall (R ~ 0.65). During MAM and JJAS the multiple regression of rainfall and lightning are very strong over all the regions (Fig. 3g,h).

A similar teleconnection between lightning and global SST over India, a sub-set of global lightning hotspot regions is also investigated to find the linkages between lightning flash density and global predictors. The mean lightning flash count averaged over Region 1 and Region 2 are correlated with global SST (2m Temperature) over the ocean (land) during MAM and JJAS (Figs. 4a1-d1, and a2-d2). The results show a strong and the most widespread relationship to lightning activity with the different modes of variability. The strongest positive correlation in the MAM is observed in the NINO4 region followed by NINO3.4 index for Region 1 (Fig. 4a1). On the other hand, the relationship for Region 1 is rather weak (not significant) over the NINO4 or NINO3.4 during JJAS, but a significant negative correlation is noted over the NINO1+2 indices (Fig. 4b1). Similarly, the CTT averaged over Region 1 is correlated with SST (Fig. 4c1,d1) during MAM and JJAS to demonstrate the role of deep convection over the regions. On the other hand, the strong negative correlation pattern is also noticed in lightning and CTT over Region 1 and Region 2 as it is well known that lightning flashes increase when CTT decreases (Price and Rind, 1992; Liu et al., 2012). The stronger correlation in lightning (Fig. 4a1) and CTT (Fig. 4c1) during MAM indicates the strongest relationship of large-scale global predictors and seasonal mean lightning flash density. The pre-monsoon CTT over Region 2 has a similar relationship (Fig. 4c2) as seen in Region 1 (Fig. 4c1). The results also pinpoint that during pre-monsoon seasons (MAM), lightning activity increases in associated with the El-Nino condition over Region 1. Interestingly, the JJAS (MAM) mean climatology of lightning flash density is stronger over Region 2 (Region 1) as already seen in figure 1. Similarly, during JJAS, CTT over Region 2 depict a strongest negative correlation in the NINO index (Fig. 4b2). We have calculated lightning flash density for the El-Nino and La-Nina composite years for MAM and JJAS seasons (Fig. 4e,f). The difference between El-Nino and La-Nina (El-Nino minus La-Nina) are shown in figure 4e and figure 4f for MAM and JJAS respectively. These results clearly demonstrate that lightning flash counts lead to an increase over Region 1 during MAM in the El-Nino condition. On the other hand, lightning flash density increases during JJAS in La-Nina conditions over Region 2. The results are valuable in the perspective of seasonal forecasting of lightning for the preparedness in advance.

The synoptic variance (period less than 10 days) of lightning flash counts averaged over study Region 1 is correlated with SST over ocean during MAM and JJAS (Fig. S1). The correlation of mean lightning flash counts averaged over different study regions with global rainfall during MAM and JJAS are also presented in Figure S2. It is interesting to note that lightning over Region 1 (Region 2) is positively (negatively) correlated with rainfall over Nino region during MAM. The results also demonstrate that there is a high predictability of lightning one season in advance, which is comparable to that of seasonal rainfall.

## 4. Summary and Conclusions

We have explored the possibility of the statistics of lightning and thunderstorm activity over the Indian subcontinent in this present study. Generally, for the seasonal predictions we do not focus on forecasts of specific days instead look for general trends/tendencies/possibilities over specific regions for a season ahead, which may helps to give warnings/alerts for that area (regions) well in advance. In particular, the extra-tropical SST influences and the strong association of the AMO with MAM and JJAS lightning flash density are equally crucial source of predictability along with ENSO. The sub-seasonal variances of lightning flash density with global SST reveals a strong positive correlation between them. This again corroborates that lightning flash density over the Indian subcontinent (Region 1) is tied to the slow varying predictable component (e.g. SST). Hence, a reliable seasonal forecasting of lightning flashes over India as well as over other global hotspot region is achievable as the mean and sub-seasonal variances of lightning are found highly linked with global predictors.

This opens up a new avenue for the most difficult and essential atmospheric hazards and would be helpful to save lives, economy of the populous country like India. Here, we have focused on the two global lightning hotspot regions (north America and Africa) and as a sub-set of global tropics specific interest over the Indian subcontinent to demonstrate the feasibility of seasonal forecasting of lightning with proper scientific basis. The multiple regression analysis demonstrate that seasonal lightning is potentially predictable all four regions of the global tropics. Over east India lightning is predictable of ~56% (R ~ 0.75) during pre-monsoon and an average of ~62% (R ~ 0.79) over all India for monsoon. The multiple correlation of rainfall is also high ~74% (R ~ 0.86) over all India for pre-monsoon along with other regions of global tropics. The different modes of variability and its linage with lightning flashes are examined by doing of lightning flash count (averaged over study region) with different indices for two seasons (pre-monsoon, MAM and Monsoon, JJAS). The new research will motivate researcher to pick up the challenging problem with dynamical model and lightning parameterizations for the lightning/thunderstorm activity forecast on the seasonal timescale.

**Acknowledgments**

We thank MoES, the Government of India, and Director IITM for all the support to carry out this work. We also duly acknowledge the HadiSST (https://www.metoffice.gov.uk/hadobs/hadisst/), the MERRA2 (https://gmao.gsfc.nasa.gov/reanalysis/MERRA-2/), the Tropical Rainfall Measuring Mission (TRMM) LIS/OTD, GPCP data sets used here. Authors duly acknowledge the National Crime Records Bureau (NCRB), Govt. of India (https://ncrb.gov.in/) for the data of total fatalities by natural disasters along with death due to lightning over India available from the "Accidental Deaths & Suicides in India (ADSI) Report", Govt. of India published by for the years 2002 to 2019 ($36^{th}$ to $53^{rd}$ series), CHAPTER – 1 (https://ncrb.gov.in/en/adsi-reports-of-previous-years). We also thank the freely available software viz. The Grid Analysis and Display System (GrADS), NCAR Command Language (NCL), Ferret-NOAA, Climate Data Operators (CDO), and Origin Lab.

**List of Figures:**

**Figure 1:** Seasonal climatology of Lightning Flashes (count/km2/year) over global tropics (Indian Region) during MAM (a(b)) and JJAS (c(d)) during 1997-2013. Different regions are marked in the figure (East and north-East India, Region 1: 82 ºE - 94 ºE; 20 ºN - 28 ºN; North-West India, Region 2: 68 ºE - 80 ºE; 25 ºN - 38 ºN; North America Region 3: 110 ºW - 71 ºW; 5 ºN - 43 ºN) and Africa (Region 4: 8 ºE - 35 ºE; 12 ºS - 15 ºN). (e) Year-wise count of Total deaths due to natural disasters (black line), death due to lightning only (blue line) over India; percentage share of death due to lightning is also shown in the red line.

**Figure 2:** Seasonal mean of total lightning flashes (Count/km$^2$/year) averaged over Region 1 and Region 2 for each year from 1997-2013 during (a) MAM (b) JJAS. El-Niño (La-Nina) years are marked as stars in red (indigo). The interannual variability in terms of anomaly (% of mean) of lightning flashes averaged over Region 1 and Region 2 during (c) MAM and (d) JJAS are also shown.

**Figure 3:** Correlation of mean Lightning Flash Count (Cloud Top Temperature; unit: K) averaged over two selected global lightning hot spot regions, Region 3 (a3, b3, c3, d3) and Region 4 (a4, b4, c4, d4) with Global SST during MAM and JJAS. (e, f): The correlation coefficients of lightning averaged over different study regions with SST over different boxes (Nino 1+2, Nino 3, Nino 3.4, Nino 4, NAO, AMO, PDO, Extra Tropics) across the globe for MAM (e) and JJAS (f) season; (g, h): The multiple regression "R value" for lightning and rainfall all regions (Region 1 to Region 4) and all India (68 ºE - 98 ºE; 8 ºN - 38 ºN) are shown for MAM and JJAS. The dashed line shows 90% significance level.

**Figure 4:** Correlation of mean Lightning Flash Count (Cloud Top Temperature; unit: K) averaged over region 1 with Global SST during MAM (a1(c1)) and JJAS (b1(d1)).In (a1) and (b1) averaged Lightning is also correlated with 2-m air temperature over land regions. a2 (c2), b2(d2) are same but for region 2. Correlation values greater than 95%(two-tailed) are stippled. Difference between Lightning flashes (count/km2/year) of El Niño and La Niña composite years for MAM (e) and JJAS (f) season.

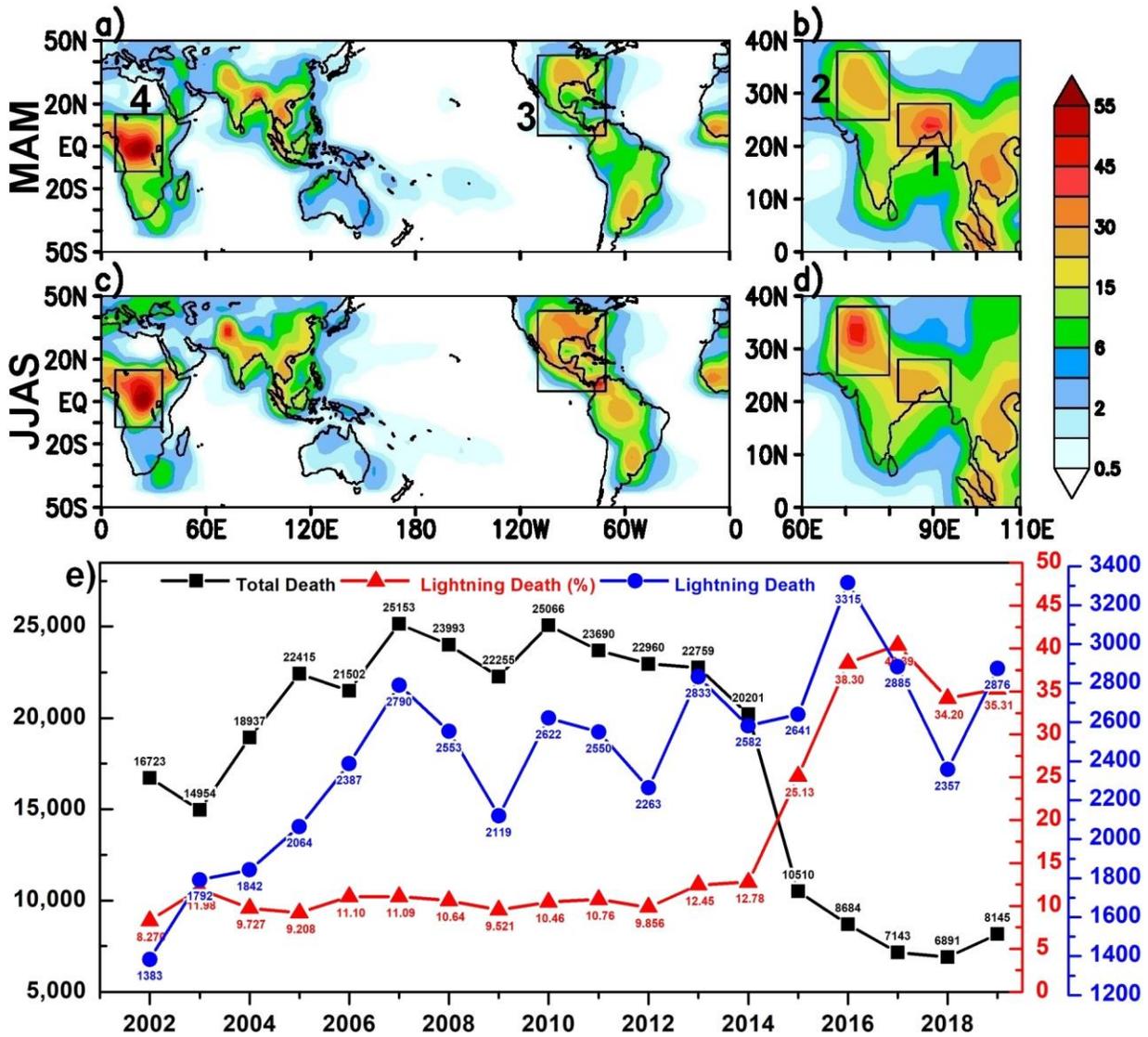

**Figure 1:** Seasonal climatology of Lightning Flashes (count/km2/year) over global tropics (Indian Region) during MAM (a(b)) and JJAS (c(d)) during 1997-2013. Different regions are marked in the figure (East and north-East India, Region 1: 82 ºE - 94 ºE; 20 ºN - 28 ºN; North-West India, Region 2: 68 ºE - 80 ºE; 25 ºN - 38 ºN; North America Region 3: 110 ºW - 71 ºW; 5 ºN - 43 ºN) and Africa (Region 4: 8 ºE - 35 ºE; 12 ºS - 15 ºN). (e) Year-wise count of Total deaths due to natural disasters (black line), death due to lightning only (blue line) over India; percentage share of death due to lightning is also shown in the red line.

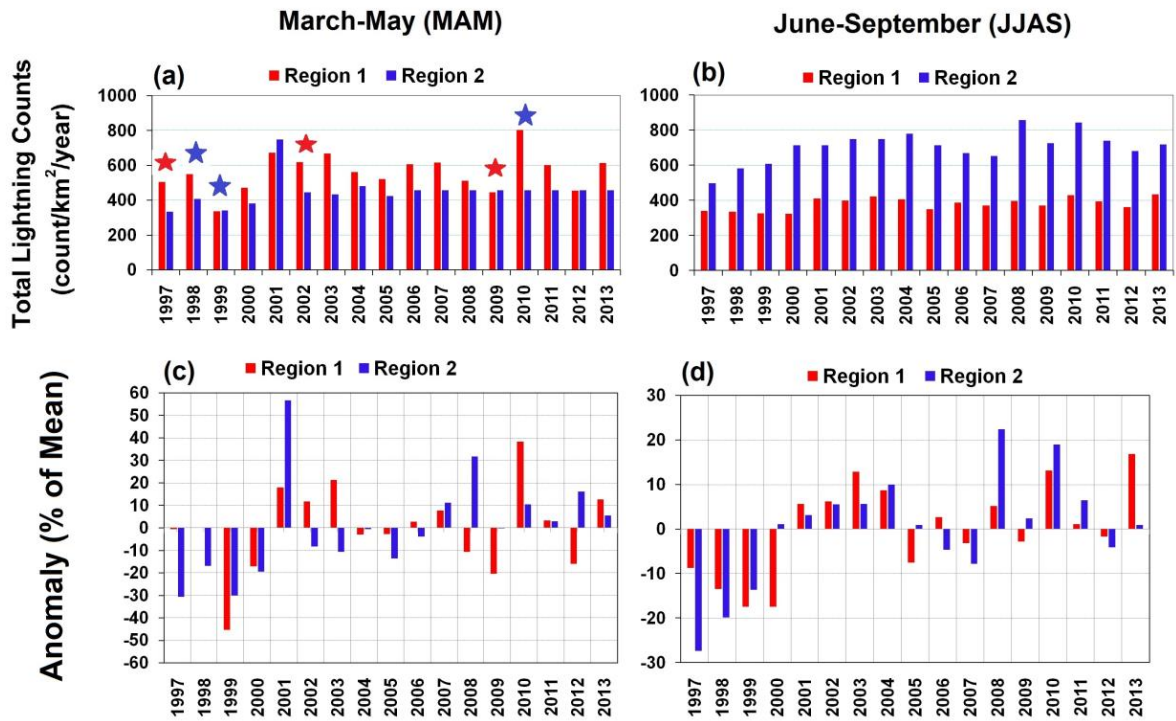

**Figure 2:** Seasonal mean of total lightning flashes (Count/km$^2$/year) averaged over Region 1 and Region 2 for each year from 1997-2013 during (a) MAM (b) JJAS. El-Niño (La-Nina) years are marked as stars in red (indigo). The interannual variability in terms of anomaly (% of mean) of lightning flashes averaged over Region 1 and Region 2 during (c) MAM and (d) JJAS are also shown.

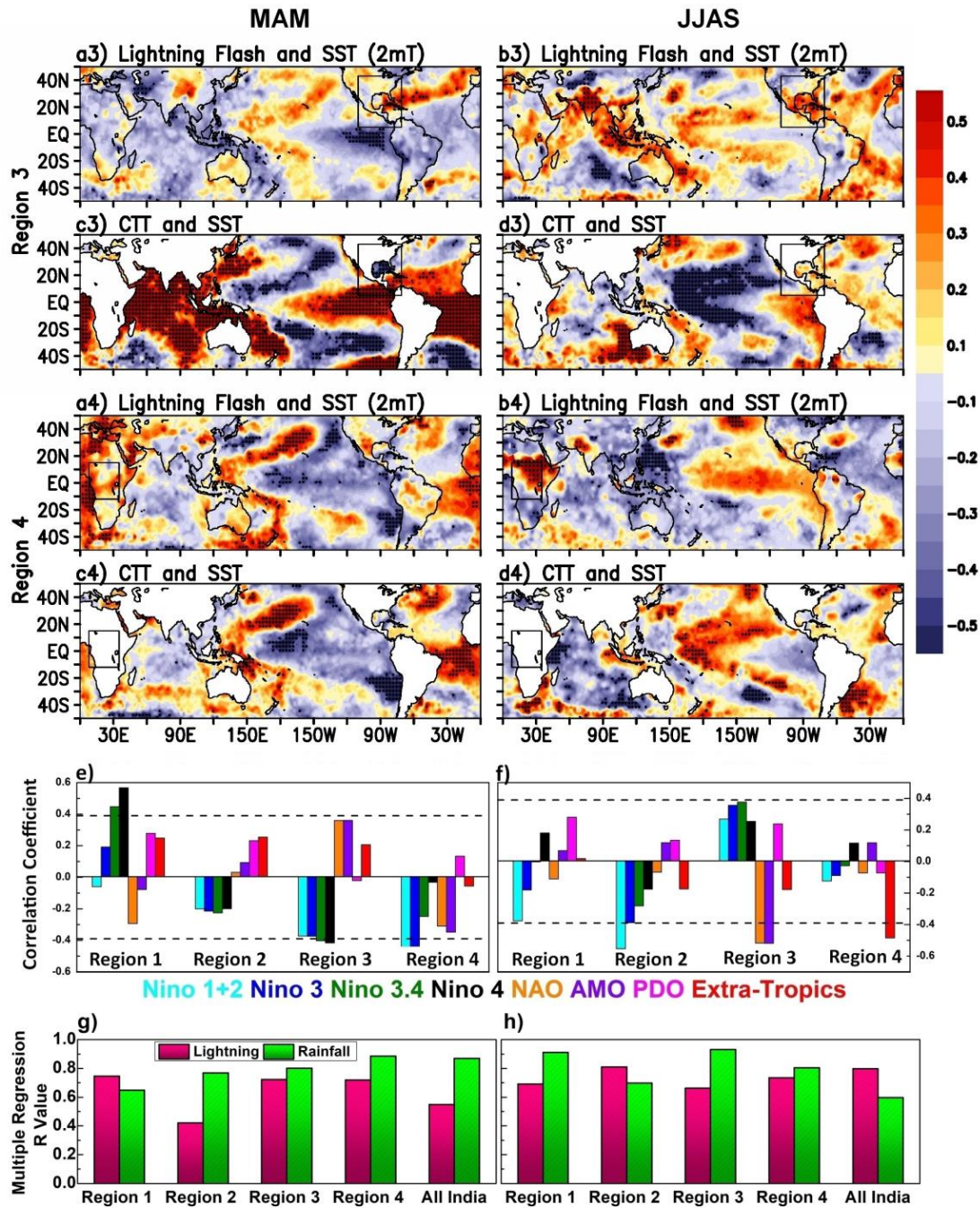

**Figure 3:** Correlation of mean Lightning Flash Count (Cloud Top Temperature; unit: K) averaged over two selected global lightning hot spot regions, Region 3 (a3, b3, c3, d3) and Region 4 (a4, b4, c4, d4) with Global SST during MAM and JJAS. (e, f): The correlation coefficients of lightning averaged over different study regions with SST over different boxes (Nino 1+2, Nino 3, Nino 3.4, Nino 4, NAO, AMO, PDO, Extra Tropics) across the globe for MAM (e) and JJAS (f) season; (g, h): The multiple regression "R value" for lightning and rainfall all regions (Region 1 to Region 4) and all India (68 ºE - 98 ºE; 8 ºN - 38 ºN) are shown for MAM and JJAS. The dashed line shows 90% significance level.

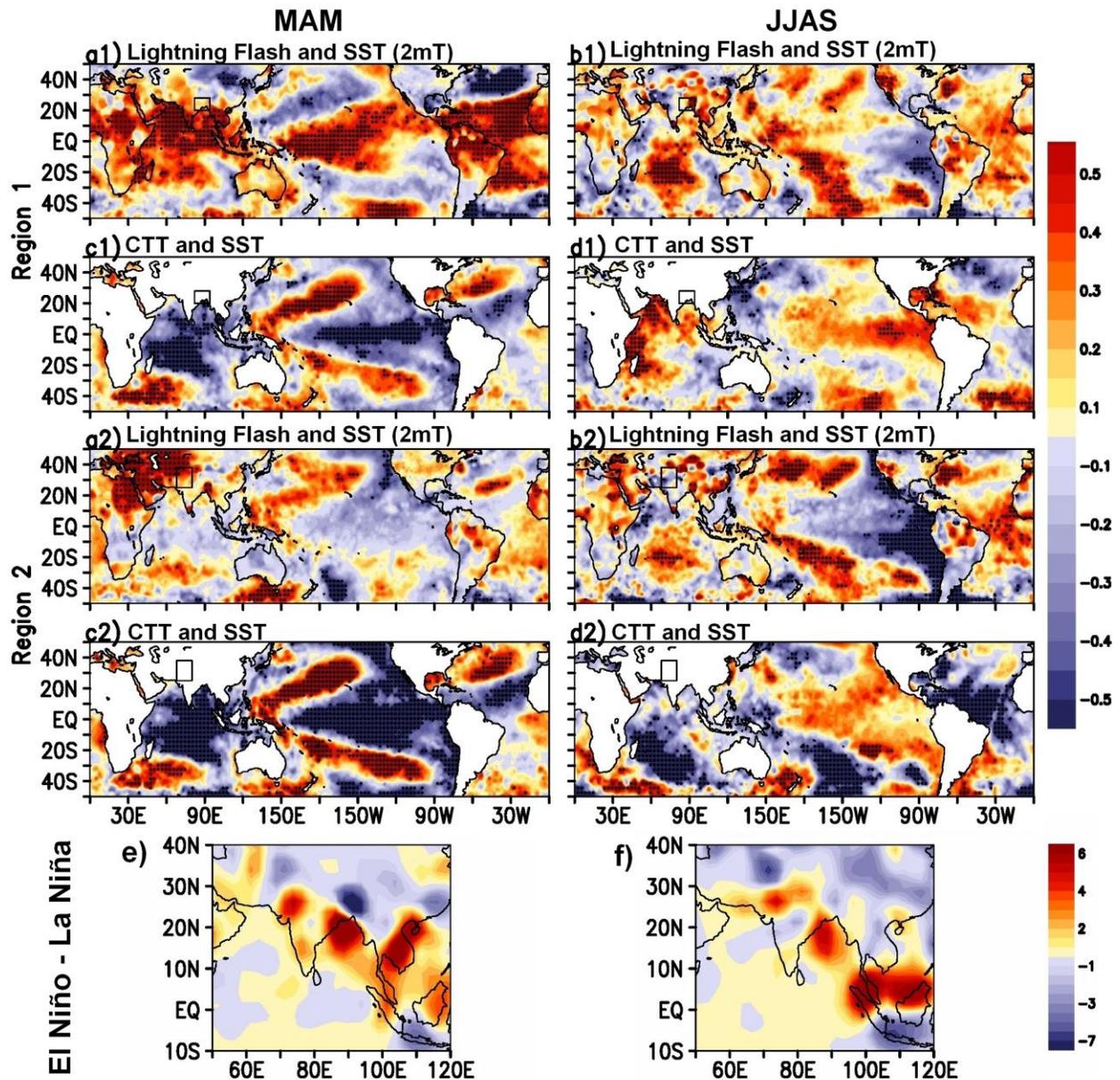

**Figure 4:** Correlation of mean Lightning Flash Count (Cloud Top Temperature; unit: K) averaged over region 1 with Global SST during MAM (a1(c1)) and JJAS (b1(d1)). In (a1) and (b1) averaged Lightning is also correlated with 2-m air temperature over land regions. a2 (c2), b2(d2) are same but for region 2. Correlation values greater than 95% (two-tailed) are stippled. Difference between Lightning flashes (count/km$^2$/year) of El Niño and La Niña composite years for MAM (e) and JJAS (f) season.

**Supplementary Information :**

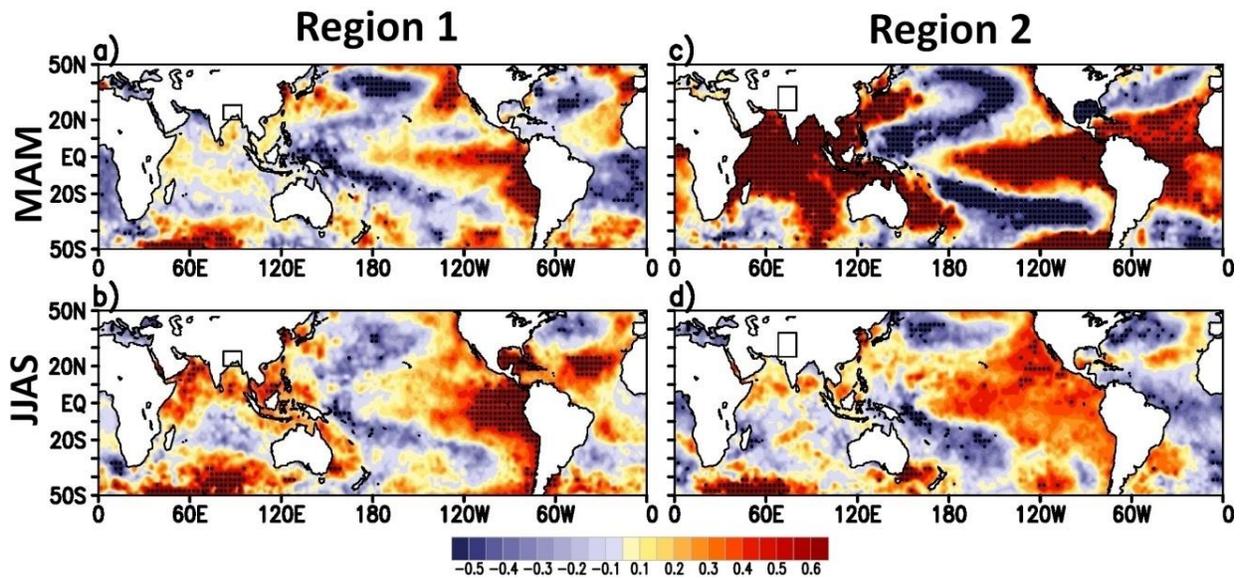

**Figure S1:** Correlation between synoptic variance (period less than 10 days) of Lightning Flash Count averaged over region 1 (region 2), and Global SST during MAM (a(c)) and JJAS (b (d)). Correlation values greater than 95% (two-tailed) are stippled.

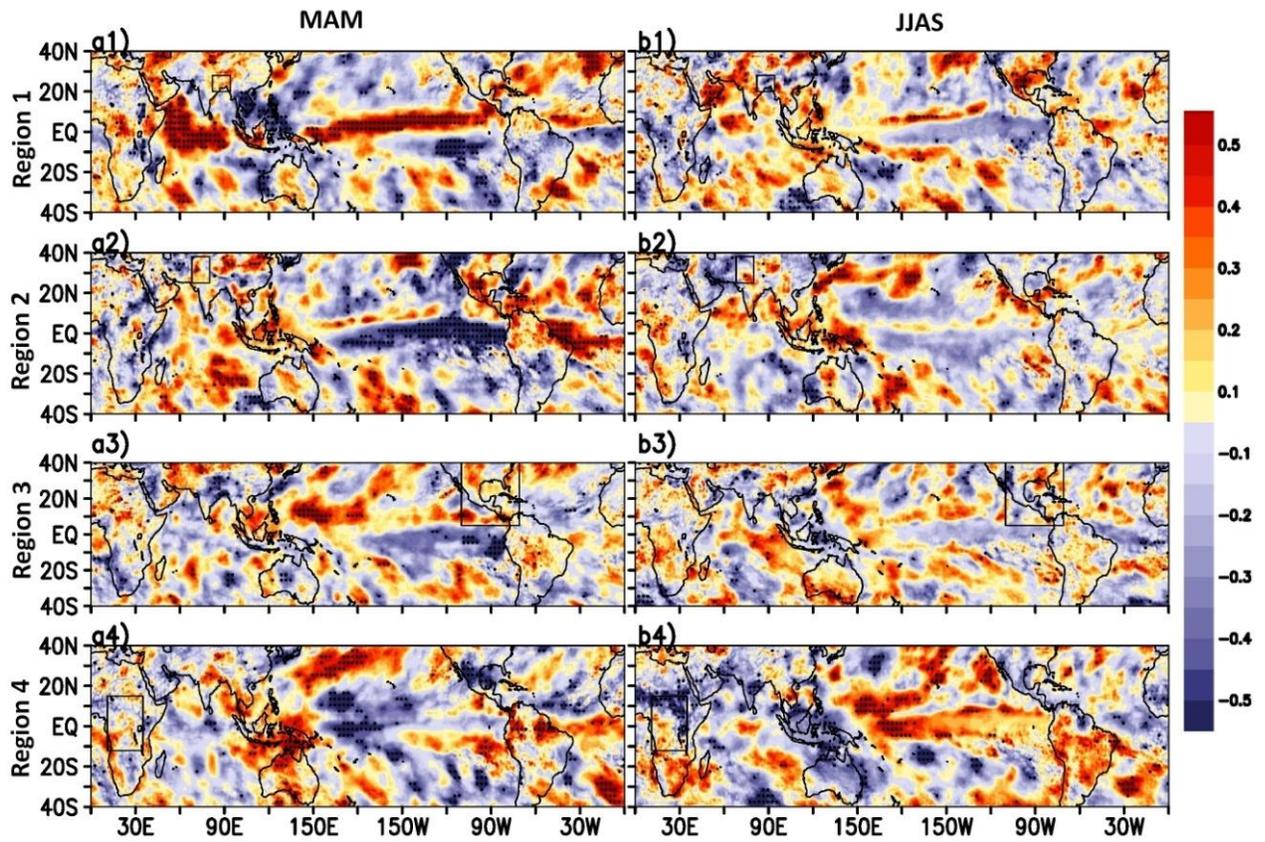

**Figure S2:** Correlation of mean Lightning Flash Count averaged over study regions with global rainfall during MAM and JJAS. Correlation values greater than 95% (two-tailed) are stippled. Unit of Lightning flash count: count/km$^2$/day.